\documentclass[twocolumn,twoside,slac_two]{revtex4}
\usepackage{graphicx}
\usepackage{fancyhdr}
\begin{document}

\title{The OPERA experiment: on the way to the direct observation of $\nu_\mu \rightarrow \nu_\tau$ oscillation}

%

\author{M. De Serio for the OPERA Collaboration}
\affiliation{Istituto Nazionale di Fisica Nucleare - Sezione di Bari, via E. Orabona 4, I-70126 Bari, Italy}

\begin{abstract}

OPERA (\emph{O}scillation \emph{P}roject with \emph{E}mulsion t\emph{R}acking \emph{A}pparatus) is a long-baseline 
neutrino experiment, designed to provide the first direct proof of $\nu_\mu \rightarrow \nu_\tau$ 
oscillation in the atmospheric sector using the \emph{C}ERN \emph{N}eutrinos to \emph{G}ran \emph{S}asso (CNGS) 
$\nu_\mu$ beam. The detector, consisting of a modular target made of lead - nuclear emulsion units 
complemented by electronic trackers and muon spectrometers, has been conceived to select $\rm{\nu_\tau}$ 
charged current interactions, among all neutrino flavour events, through the observation of the outcoming tau leptons 
and subsequent decays.  
In this paper, the detector, the event analysis chain and the preliminary results from the first OPERA physics run are reported. 

\end{abstract}

\maketitle

\thispagestyle{fancy}


\section{Introduction}

Several experiments, using both natural and artificial sources, have provided strong evidence 
in favor of the hypothesis of neutrino oscillations over the past 20 years. 
Flavour conversion in the atmospheric sector, first established by the Super-Kamiokande experiment 
\cite{SK} and more recently confirmed by K2K \cite{K2K} and MINOS \cite{MINOS}, has been assessed in disappearance mode, namely in terms 
of an observed deficit of muon neutrinos reaching the detector. Nevertheless, an unambiguous proof of 
the oscillation mechanism through the direct detection of $\nu_\tau$'s from $\nu_\mu$ conversion is still 
a missing piece. 

The OPERA experiment \cite{OPERA}, currently taking data, has been designed to observe $\nu_\tau$ appearance 
in an almost pure $\nu_\mu$ beam (CNGS), produced at CERN and sent towards the Gran Sasso underground laboratory 
(LNGS, Italy), with an average rock coverage of $1400 \, \rm{m}$,  over a baseline of $730 \, \rm{km}$. 

The construction of the detector started in 2003 and was completed in 2008. After a short physics run in 
October 2007, the experiment started full data-taking in 2008, when about $1700$ neutrino interactions 
were collected. 
In the following, the first results from the 2008 run, as well as a description of the detector and 
the experimental strategy, are presented.

\section{OPERA experimental signature and $\tau$ search potential}

The CNGS beam is a wide-band $\nu_\mu$ beam with an average energy of $17 \, \rm{GeV}$, well above 
the $\tau$ production threshold, designed to maximise the number of $\rm{\nu_\tau}$ charged current 
interactions at LNGS. The prompt $\rm{\nu_\tau}$ production is negligible. 
The $\rm{\overline{\nu}_\mu}$ contamination is $\sim 4 \%$; the total $\rm{\nu_e}$ and 
$\rm{\overline{\nu}_e}$ contamination is below $1 \% \,$. 

The $\rm{\nu_\tau}$ appearance signature is the detection of the decays of the $\rm{\tau}$ leptons 
produced in $\rm{\nu_\tau}$ CC interactions in one prong (muon, electron or hadron) or in three prongs. 
Given the short $\rm{\tau}$ decay length (about $1 \, \rm{mm}$ at the average CNGS beam energy), 
nuclear emulsion films are used as tracking detectors with sub-micrometric and sub-milliradian resolutions  
in order to identify decay topologies and efficiently reject the background. 

Table \ref{Ev_Bkg} shows the numbers of identified $\rm{\nu_\tau}$ events to be collected in 5 years of
data-taking assuming nominal CNGS intensity ($4.5 \times 10^{19}$ p.o.t./year) for 
$\rm{\Delta m^2} = 2.5 \times 10^{-3} \, \rm{eV^2}$ and maximal mixing. 
For each $\rm{\tau}$ decay channel included in the analysis, the total number of expected background events 
is also reported. OPERA is expected to observe about 10 signal events with a background of less than one event.

\begin{table*}[t!]
\caption{Expected numbers of signal and background events in 5 years of run at
    nominal CNGS intensity. Maximal mixing is assumed.}\label{table:1}
\small
\begin{center}
\begin{tabular}{|l|c|c|}
\hline
\textbf{$\rm{\tau}$ decay channel} & \textbf{Signal} & \textbf{Background} \\
{} &{\textbf{$\rm{\Delta m^2} = 2.5 \times 10^{-3} \, \rm{eV^2}$}} &{}\\
\hline
$\rm{\tau} \rightarrow \rm{\mu}$                & 2.9 & 0.17 \\
{} &{} &{} \\
\hline
$\rm{\tau} \rightarrow \rm{e}$                & 3.5 & 0.17 \\
{} &{} &{} \\
\hline
$\rm{\tau} \rightarrow \rm{h}$                & 3.1 & 0.24 \\
{} &{} &{} \\
\hline
$\rm{\tau} \rightarrow 3 \, \rm{h}$                & 0.9 & 0.17 \\ 
{} &{} &{} \\
\hline
ALL & 10.4 & 0.75 \\ 
{} &{} &{} \\
\hline
\end{tabular}\\[2pt]
\label{Ev_Bkg}
\end{center}
\end{table*}

\section{The detector}

\begin{figure*}[]
\centering
\includegraphics[width=130mm]{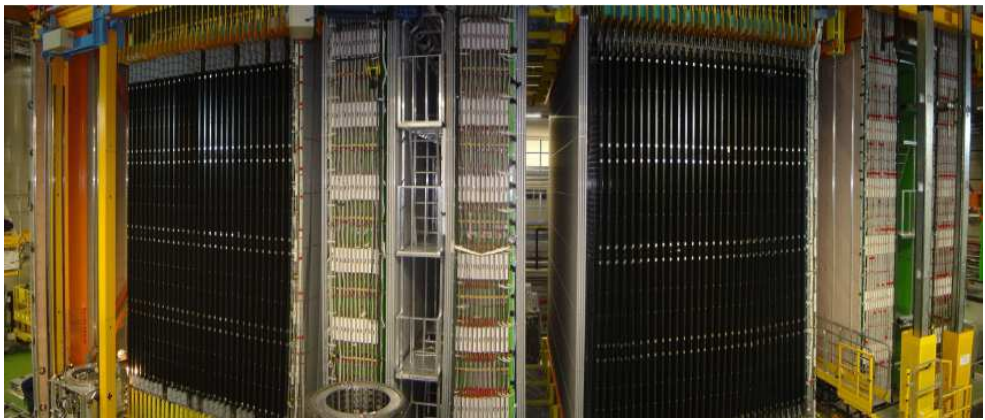}
\caption{The OPERA detector installed in the Hall C of LNGS.
    \label{fig:detector}}
\end{figure*}

OPERA is a hybrid detector made of two identical super-modules, each one consisting of a target section 
followed by a muon spectrometer (Fig. \ref{fig:detector}). 
 
The target section, with a total mass of $1.25 \, \rm{kt}$, has a modular structure and is segmented 
into basic units called \emph{bricks}. 
A brick consists of 57 thin nuclear emulsion films, acting as sub-micrometric resolution trackers,
interleaved with 56 $1 \, \rm{mm}$-thick lead plates, used as passive material. 
The brick transverse dimensions are $12.8 \times 10.2 \, \rm{cm^2}$; the thickness along the beam 
direction is about $7.9 \, \rm{cm}$, corresponding to $\sim 10$ radiation lengths. 
A box containing two additional films, called \emph{Changeable Sheets} (CS), is glued on the downstream face 
of each brick. CS films are used to validate the event signal prior to the unpacking of the brick 
and to provide predictions for the event location. 
Bricks are arranged in planar structures, called \emph{walls}, with transverse dimensions
$\sim 6.7 \times 6.7 \, \rm{m^2}$. Each wall is coupled to a pair of tracker planes (TT) 
providing bi-dimensional track information. They are made of plastic scintillator strips of 
$1 \, \rm{cm}$ thickness, $2.6 \, \rm{cm}$ width and $6.9 \, \rm{m}$ length, arranged in modules and 
readout by WLS fibers and multi-anode 64-pixel PMT's at both ends. Each tracker plane consists of 4 modules 
of horizontal or vertical strips. 
A minimum ionising particle typically produces about $6$ photo-electrons. This results in a high detection  
efficiency ($\sim 99 \%$), required to provide the event trigger signal. TT planes have the primary task 
to identify in real-time the brick where the $\rm{\nu}$ interaction occurred.

The spectrometer of each super-module is an instrumented dipolar magnet ($\sim 8.75 \times 8 \, \rm{m^2}$) 
made of two magnetized iron walls producing a field of $1.53 \, \rm{T}$ in the tracking region with vertical lines of
opposite directions in the two walls. In between the 12 iron slabs of each wall, planes of bakelite RPC's 
are inserted to measure the range of stopping particles and track penetrating muons. 
Each RPC has dimensions $2.9 \times 1.1 \, \rm{m^2}$. 
The muon charge and momentum are measured by 6 planes of drift tubes with $38 \, \rm{mm}$ diameter and $8 \, \rm{m}$ length, 
placed in front, behind and in between the magnet walls. The spatial resolution ($\sim 300 \, \rm{\mu m}$) and 
high efficiency ($> 99 \%$) result in a very low probability of wrong charge sign assignment ($< 0.3 \%$), relevant 
for background rejection, and high momentum resolution (better than $20 \%$) below $50 \, \rm{GeV} / c$. 
Ambiguities in the reconstruction of multi-track events are solved by means of two additional planes of RPC's with 
crossed readout strips ($\pm 45^\circ$), placed upstream of each dipolar magnet to complement the information 
provided by the drift tube planes. 

The muon identification efficiency, obtained by a combined analysis of spectrometer and scintillator 
tracker data, is about $95 \%$. 

A veto system, consisting of planes of glass RPC's, is placed in front of the first super-module and allows 
to tag the interactions occurring in the upstream rock.

\section{OPERA step by step}

\begin{figure*}[ht!]
\centering
\begin{tabular}{cc}
    \includegraphics[width=70mm]{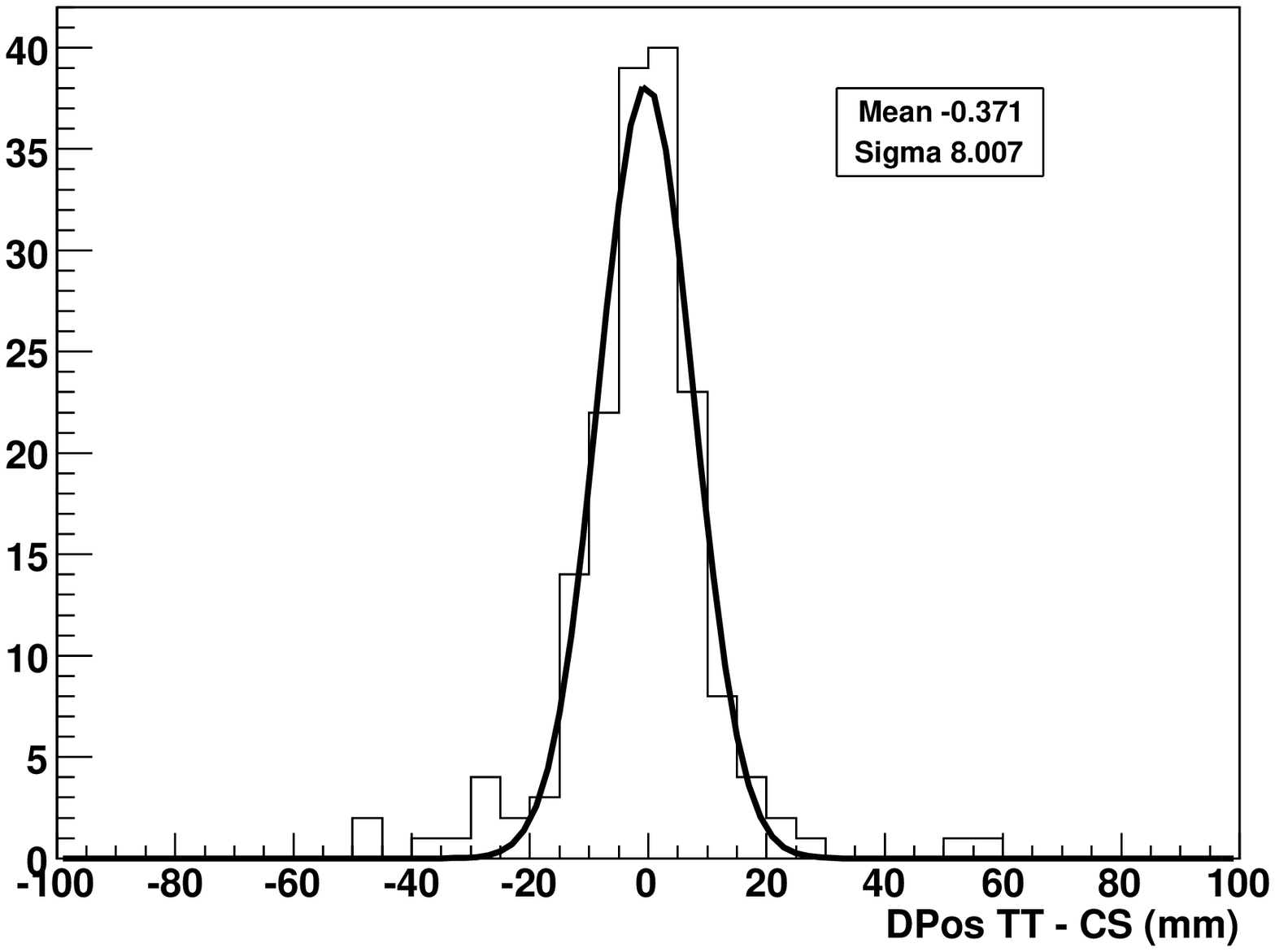} & \includegraphics[width=73mm]{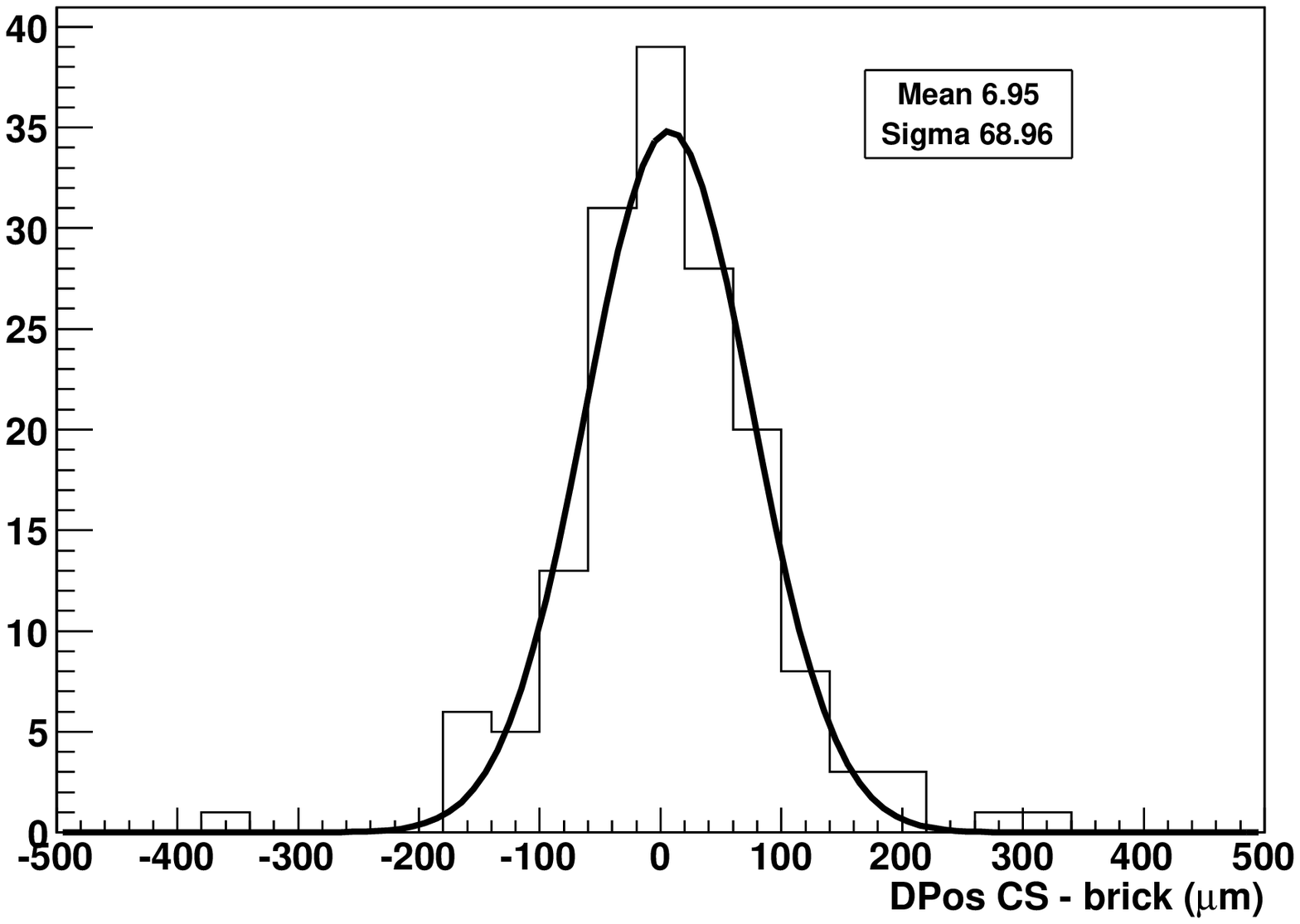}
\end{tabular}
\caption{Left: position residuals between CS measurements and TT predictions for a sample of muons. 
Right: position residuals between CS predictions and measurements in the most downstream film of the brick 
for a sample of muons.} \label{fig:TTCSBrick}
\end{figure*}

The selection of beam-related events is performed off-line, based on a time stamp 
with a synchronisation accuracy of $100 \, \rm{ns}$ between the CNGS tagging and the OPERA timing systems.  
Neutrino interactions occurring in the surrounding material are filtered out by an algorithm that 
classifies in-target events into CC and NC interactions. 
For each event, by a combined analysis of TT and spectrometer data, a map, assigning a probability 
to contain the neutrino interaction vertex to each of the bricks in the event region, is built. 
The brick with the highest probability is extracted from the target. The CS films are developed underground  
and analysed by means of high-speed automatic microscopes in order to confirm the presence of the interaction. 
If tracks compatible with the reconstruction of the electronic detectors data are found, 
the brick is brought to the surface laboratory and exposed to a controlled flux of high-energy cosmic rays, 
required for film by film alignment with sub-micrometric accuracy. 
The brick is then unpacked and the emulsion films are developed. All CS tracks are searched for 
in the most downstream film of the brick and those successfully located are followed back 
film by film up to the points where they originate. The \emph{disappearance} of a track is a signature 
of either a primary or a secondary vertex. In order to confirm the presence of a vertex and fully reconstruct 
the event, an area of about $1 \, \rm{cm^2}$ is measured in several films upstream and downstream with respect to 
the candidate vertex point(s). All tracks originating inside the volume are analysed by a vertex 
reconstruction algorithm and a decay search procedure is applied to select interesting topologies. 
A detailed kinematical analysis, including momentum measurement by Multiple Coulomb Scattering, electron 
identification and e.m. shower energy reconstruction, is finally applied to candidate events.

\section{The CNGS 2008 physics run}

After a short run in 2007, when only 38 neutrino interactions were collected in 3.6 effective days 
of running due to CNGS beam failures, the first OPERA physics run took place from June to November 
2008. An intensity of $1.78 \, \times 10^{19}$ p.o.t. was integrated over the entire run  
with an average efficiency of the CNGS complex of about $60 \, \%$. All OPERA electronic 
detectors were operational and the livetime of the data acquisition exceeded $99 \, \%$.
OPERA collected 10100 events on time and, among them, about 1700 interactions in the target.
At the time of writing, about 900 neutrino events, distributed among several laboratories in Europe and 
in Japan, have been located. 

The key role of the CS films as an interface detector between electronic trackers and target units, 
performing efficient track matching and powerful background rejection, is demonstrated in 
fig. \ref{fig:TTCSBrick}. The distribution of the residuals between the measured positions 
and the TT-predicted impact points in the CS films for muon tracks shows a sigma of less than $1 \, \rm{cm}$. 
The precision in track connection from the CS films to the brick is better than $100 \, \rm{\mu m}$, thus an improvement 
by more than two orders of magnitude is achieved.

\begin{figure}[b!]
\centering
\includegraphics[width=70mm]{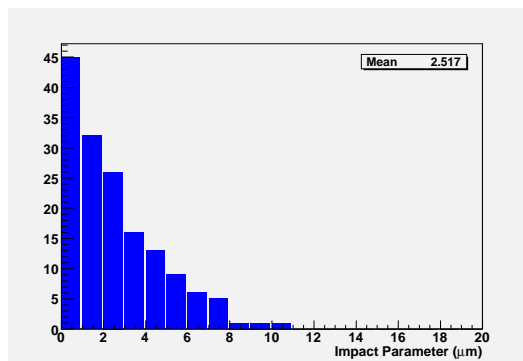}
\caption{Impact parameters of muon tracks with respect to reconstructed vertices.} \label{fig:IP}
\end{figure}

The distribution of the impact parameters of muon tracks with respect to reconstructed vertices 
is shown in fig. \ref{fig:IP}. The mean value of $2.5 \, \rm{\mu m}$ is the result of a precise inter-calibration 
of emulsion films using cosmic ray tracks to correct for relative misalignments and deformations on a 
local scale around the interaction vertex.

\begin{figure*}[]
\centering
\includegraphics[width=120mm]{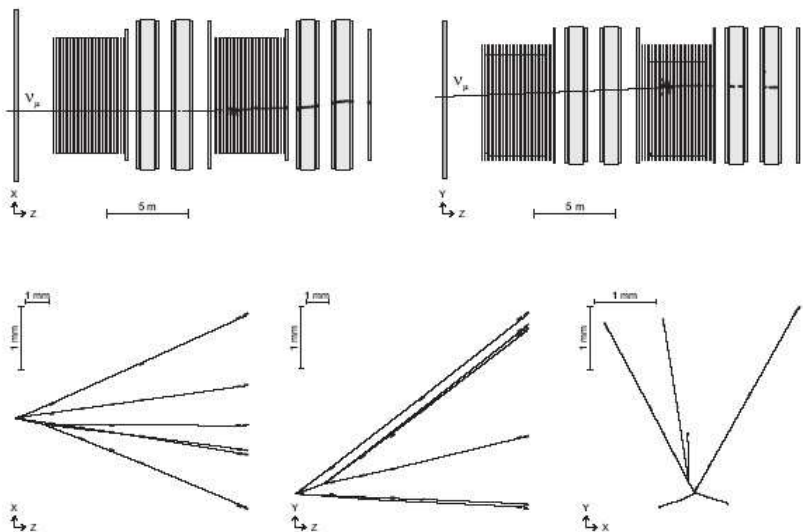}
\caption{Online display of the OPERA electronic detectors (top) and emulsion reconstruction (bottom) 
of a $\nu_\mu$ charged current interaction showing a charm-like topology.} \label{fig:Charm}
\end{figure*}

The production of charmed particles and subsequent decays play a crucial role for the experiment. 
Since charm decays have the same topology as $\tau$ decays, their detection represents a valid 
cross-check of the $\tau$ identification efficiency. Fig. \ref{fig:Charm} shows an event 
with a 4-prong primary vertex, originating at a depth of about $30 \, \rm{\mu m}$ in the upstream lead plate, 
and a 3-prong secondary vertex with a charged parent track produced in the interaction, located 
at a distance of $1150 \, \rm{\mu m}$ from the neutrino vertex point. 
The muon and the candidate charmed particle lie in a back-to-back configuration ($\Delta \phi 
\sim 150^{\circ}$), as expected in the case of charm production.  
The probability of a hadron interaction has been evaluated to be about $10^{-6}$, the probability 
of a decay in flight of a K is about $10^{-3}$.

\section{Conclusions} 

The OPERA experiment is taking data from the CNGS beam. 
During the first physics run in 2008, an integrated intensity of $1.78 \, \times \,  10^{19}$ p.o.t. 
was collected. The analysis of 2008 data is in progress, as well as a full re-evaluation 
of the efficiencies and the background. 
The observation of several charm decay candidates proves the capability of OPERA to detect 
short-lived particle decay topologies. The concept of the experiment has been fully validated. 

In 2009 run, an integrated intensity of $3.2 \, \times \,  10^{19}$ p.o.t. is foreseen. At the time of writing, 
$1.93 \, \times \,  10^{19}$ protons have already been collected with an average CNGS efficiency of $74 \, \%$. 

About 2 $\tau$'s are expected in the 2008 - 2009 data sample. 

The OPERA experiment is on the way to the direct observation of $\nu_\mu \rightarrow \nu_\tau$ oscillation. 

\bigskip 

\end{document}